# Probing the Extraordinary Ends of Ordinary Stars: White Dwarf Seismology with the Whole Earth Telescope

Steven D. Kawaler
*Department of Physics and Astronomy*
*Iowa State University*
*Ames, IA 50011 USA*

**Abstract.** During the final evolution of most stars, they shed their outer skin and expose their core of the hot ashes of nuclear burning. As these hot and very dense cores cool into white dwarf stars, they go through episodes of multiperiodic, nonradial g-mode pulsation. The tools of stellar seismology allow us to use the pulsation spectra as powerful probes into the deep interiors of these stars. Progress in white dwarf seismology has required significant international cooperation, since another consequence of the complex pulsations of these stars is decoding the true pulsation frequencies requires a coordinated global effort involving high-speed photometric observations. Through one such effort, the Whole Earth Telescope project, we have located subsurface composition changes, detected differential rotation and magnetic fields, and measured fundamental quantities such as stellar mass, luminosity, and distance to extraordinary accuracy.

## 1. Introduction

Models of the evolution of young hot white dwarfs are incompletely constrained by standard techniques of spectroscopy and broad band photometry. A new and powerful tool for probing the depths of these stars is "temporal spectroscopy". This observational tool of asteroseismology can be applied to those objects that show periodic photometric variability. Fortunately, nonradial *g*-mode pulsations are observed in white dwarf stars during three phases of their evolution. Through temporal spectroscopy, we have accurately determined of the mass, rotation rate, cooling/contraction rate, compositional stratification, and magnetic field strength of many white dwarfs over a range of ages.

The phases of white dwarf evolution in which some are observed to pulsate span the time from their birth to middle age. During the hot phases of white dwarf evolution ($T_\text{eff} > 80,000$ K), some hydrogen–deficient white dwarfs that show carbon and oxygen enrichment are observed to pulsate with periods of 300–1000 seconds. Along with these "DOV" stars, several planetary nebula nuclei with similar C and O enrichment show variations, with periods of 1000–2000 seconds (Bond, Ciardullo, & Kawaler 1993). At cooler effective temperatures (near about 25,000 K) lie the DBV stars, pulsating white dwarfs with helium–dominated spectra. Finally, in a narrow strip near $T_\text{eff} \approx 12,000$ K we find the DAV (or ZZ Ceti) stars, which are pulsating white dwarfs with hydrogen dominated spectra. Most, if not all white dwarfs in the DAV instability strip are variables (Winget 1988, Bergeron et al. 1995).

Several recent reviews of white dwarf pulsation (i.e. Winget 1988, Kawaler & Hansen 1989, Kawaler 1991) discuss white dwarf seismology as "showing significant promise" as a technique for constraining stellar models. With the success of the Whole Earth Telescope (Nather et al. 1990) in recent years, this promise has become realized. Temporal spectroscopy of white dwarfs addresses many important questions about white dwarf evolution. By measuring the surface compositional stratification, seismological observations directly address the formation of planetary nebula nuclei, the evolutionary status of the PG 1159 and DB stars, and the chemical evolution of white dwarfs of all spectral types. In addition, by measuring the thickness of the surface helium and hydrogen layers, temporal spectroscopy of white dwarf stars has a direct impact on use of the white dwarf luminosity function as a probe of the star formation and age of the galactic disk (Wood 1992). This paper reviews examples of the results of Whole Earth Telescope analyses of DO and DB white dwarfs, and their impact on current models of white dwarf evolution.

## 2. Nonradial Pulsation Tools

The equations describing nonradial pulsations of a star are fairly complex. However, their asymptotic limits compactly describe the observed pulsation periods of white dwarf stars. Here I present these relations in the form of seismological "tools" that we use in analysis of the pulsation spectra of white dwarfs. We talk of pulsation *modes* as having frequencies $\nu_{n\ell m}$. A given mode has $\ell$ nodal lines on its surface, of which $m$ nodal lines pass through the axis of symmetry; $n$ is the number of nodes in the radial direction. The asymptotic analysis provides expressions for the frequency $\nu_{n\ell}$, or period $\Pi_{n\ell}$, in terms of $n$, $\ell$, and integral properties of the equilibrium model. The periods of nonradial modes do not depend on $m$ unless rotation (or another physical process) disturbs the spherical symmetry of the star. In that case, mode frequencies split into $2\ell + 1$ components — assuming all $m$ components are indeed present in the star. In this section we outline how to use period spacings and frequency spacings as probes of the innards of white dwarfs.

### 2.1. $g$-mode Period Spacings

The asymptotic limit at low frequency (smaller than the radial fundamental pulsation frequency) corresponds to $g$-modes, where buoyancy is the dominant restoring force. For $g$-modes, the pulsation modes are more-or-less equally spaced in period. The periods $\Pi_{n\ell}$ (in seconds) are given by

$$\Pi_{n\ell} = \frac{\Pi_o}{\sqrt{\ell(\ell+1)}}(n + \epsilon), \qquad (1)$$

where $\Pi_o$ is a constant that depends only on the the equilibrium structure of the star. Therefore $g$-modes with consecutive values of $n$ are equally spaced in period for modes with the same value of $\ell$. The period spacing for $\ell = 1$ modes should be a factor of $\sqrt{3}$ times larger than for $\ell = 2$ modes in the same star.

This gives us the first of our seismological diagnostics. Identification of uniform period spacing in a pulsating star gives us $\Pi_o$ and $\ell$, placing immediate and important constraints on the stellar properties. Kawaler (1988) and Kawaler and Bradley (1994) show that, for DO white dwarfs, $\Pi_o$ depends primarily on the mass of the star. For PG 1159 models between $0.4M_\odot$ and $0.8M_\odot$ and $L \geq 10L_\odot$,

$$\Pi_o = 15.5 \left(\frac{M}{M_\odot}\right)^{-1.3} \left(\frac{L}{100L_\odot}\right)^{0.035} \tag{2}$$

with very weak dependence on the thickness of the surface helium layer.

In cooler white dwarfs, where the nondegenerate surface layers determine the pulsation periods, $\Pi_o$ does depend on $T_{\text{eff}}$ (i.e. Kawaler 1987, Tassoul, Fontaine, & Winget 1990); however, the instability strips within which the pulsating stars are found are very narrow in temperature, and thus the period spacing is more-or-less dependent again only on mass (Bradley & Winget 1994).

## 2.2. Departures From Uniform Spacing: Mode Trapping

Equation (1) holds for idealized stellar models with homogeneous composition. Real white dwarfs are compositionally stratified. The steep subsurface composition gradients produce steep gradients in the density that can act as internal reflecting boundaries for oscillations. This leads to the phenomenon of "mode trapping": modes that have nodes within the composition transition zones can have the amplitude of their eigenfunctions greatly reduced in the interior. The eigenfunctions are, in effect, pinched off below the surface at the composition transition zones. Because trapped modes involve oscillation of a small fraction of the total mass of a star, they require less energy to excite to observable amplitude than untrapped modes. Winget, Van Horn, & Hansen (1981) showed that this property of trapped modes provides a natural mechanism for the preferential selection of trapped modes in ZZ Ceti stars.

More recent work demonstrates how identification of trapped modes allows us to determine the thickness of the surface layers of white dwarfs. The method follows from an analytic analysis by Carl Hansen (1987, private communication). Numerical work confirmed the basic results of Hansen's analysis (see Kawaler & Weiss 1990 and Brassard et al. 1992a). The key element is that trapped modes show departures from uniform period spacing. The period difference between adjacent (in $n$) modes is the principal observational diagnostic of mode trapping. Generally, trapped modes are closer in period to the next lowest–order mode than the mean spacing. Once identified, the periods of the trapped modes allow determination of the depth of composition transition zones; thin surface layers show trapped modes that are much farther apart in period than thick suface layers. An additional result is that the periods of trapped modes of the same $\ell$ should show fixed ratios that are largely independent of the stellar parameters.

Thus the second tool is the period spacing diagram: a plot of $\Delta \Pi$ versus $\Pi$. Minima in $\Delta \Pi$ usually correspond directly to trapped modes. Once trapped modes are identified, the Hansen relation, along with more detailed models, determines the position below the surface at which a mode sees a reflective boundary. For many stars that have been observed with the Whole Earth Telescope, trapped modes have been identified either from minima in the period spacing diagram or from comparison with the trapping period ratios. For more details about the phenomenon of mode trapping and applications to DA, DB, and DOV models see Kawaler & Weiss (1990), Kawaler (1991), Bradley & Winget (1991), Brassard et al. (1992a,b), Bradley, Winget & Wood (1993), Kawaler & Bradley (1994), and Bradley & Winget (1994).

### 2.3. Rotational Splitting of $g$-modes

In nonrotating, axially symmetric stars, the eigenfrequency does not depend on the azimuthal quantum number $m$. Hence the frequencies of modes with the same $\ell$ and $n$, but different values of $m$ are identical in such stars. Departures from spherical symmetry will break this $m$ degeneracy, and lead to modes with different $m$ having different frequencies. For slow ($P_{rot} \gg \Pi_{n\ell}$) uniform rotation,

$$\nu_{n\ell m} = \nu_{n\ell} - \frac{m}{P_{rot}}(1 - C_{n\ell}) \qquad (3)$$

where $\nu_{n\ell}$ is the frequency of a mode for the nonrotating case, $\nu_{n\ell m}$ is the observed frequency, and $C_{n\ell}$ is a number that is a function of the perturbation eigenfunctions. For white dwarfs, where the amplitude of oscillatory motion in the horizontal direction is much larger than in the vertical, $C_{n\ell} \simeq [\ell(\ell + 1)]^{-1}$ (Brickhill 1975).

Slow uniform rotation causes splitting of pulsation frequencies into $m$ components (assuming modes with all available values of $m$ are excited) that are equally spaced in frequency. Since for a given $\ell$ there are $2\ell + 1$ possible values of $m$, this means that $\ell = 1$ modes appear as triplets in a power spectrum, and $\ell = 2$ modes appear as quintuplets. The splitting of these multiplets will be different for different values of $\ell$, with $\ell = 1$ triplet splittings being 0.6 times the splitting of $\ell = 2$ quintuplets.

### 2.4. The Whole Earth Telescope

The above effects result in very complex light curve for a star that pulsates in only a few modes. If more than one mode is present in a star, then those modes will beat against one another. The light curve will show variations on time scales of the individual modes and on time scales corresponding to the frequency *differences* between the modes. The resulting light curves are extremely difficult, if not impossible, to fully decode from a single observing site. This is true of most nonradial pulsators: the large number of pulsation modes leads to beating effects with frequencies on the order of 1 cycle per day. Because a single observing site cannot observe its target when clouds, trees, or the Earth gets in the way, diurnal aliases necessarily appear in the Fourier power spectra of the data, and can hide true features of the star's power spectrum.

To solve this problem requires observations that are free from the 1 cycle per day aliases, and the only way to do that is to observe them 24 hours per day. This necessitates coordinated observations at observatories at widely distributed longitudes. Such observations are obtained for pulsating white dwarfs using an extremely remarkable instrument, the Whole Earth Telescope ("WET"). As described in detail by Nather et al. (1990), the WET is a global network of telescopes designed for highly coordinated interactive photometric observations of variable stars. The WET has demonstrated itself to be an extremely powerful instrument; one that has obtained data of extremely high quality on several pulsating white dwarfs. WET data allow us to use the tools of stellar seismology at nearly their full potential.

## 3. Selected Results for Hot White Dwarfs

The PG 1159 stars are hot and luminous pre–white dwarfs. The designation corresponds to the spectral class characterized by CIV and OVI spectral lines with emission cores; see the paper by Werner in these proceedings, and references therein, for more details. Central stars of some planetary nebulae show spectra very similar to the "naked" PG 1159 stars. As the precursors of more ordinary white dwarfs, the PG 1159 stars are the transition objects between stars with active nuclear burning (AGB stars) and stars that are destined to simply cool and fade. Probing their interior structure directly constrains the transition from AGB to white dwarf, and the properties of both classes of stars.

PG 1159-035 was discovered to be a variable star by McGraw et al. (1979), and is the prototype of the DOV class of pulsating white dwarfs. McGraw et al. (1979) immediately recognized the astrophysical importance of these objects: they are rapidly evolving from the planetary nebula phase to the white dwarf cooling track. Thus they are among the hottest, and most rapidly evolving, stars in the Galaxy. NLTE model atmosphere analysis by Werner, Heber, & Hunger (1991) shows that its effective temperature is close to $140,000K$; the effective temperatures of the PG 1159 spectral class ranges from $80,000K$ to $160,000K$.

### 3.1. PG 1159-035

Through temporal spectroscopy with the WET, the prototype of the pulsating hot white dwarfs, PG 1159, has yielded a wealth of information about what one of these stars look like on the inside. The reports of these observations and their analysis (Winget et al. 1991, Kawaler & Bradley 1994) show what such stars can teach us. PG 1159 is wonderful in that it shows all that we expect to see (and more) in these stars. An unbroken sequence of dozens of periodicities match very nicely with the expectations from stellar evolution and pulsation theory.

The noise level in the region of greatest power is less than $7 \times 10^{-4}$ relative amplitude. This sensitivity allows identification of much lower amplitude peaks in the power spectrum that had been possible with single–site data; Winget et al. (1991) completely resolve all of the peaks (or, more precisely, bands of power) present in earlier data into groups of individual peaks. Some of these groups are triplets of individual peaks separated equally in frequency. Other groups show 4 or 5 component peaks, also equally spaced in frequency. The frequency spacing between components in the triplets is 4.2 $\mu$Hz, while the spacing in the quintuplets is 7.0 $\mu$Hz. The expected ratio of frequency splitting for $\ell = 1$ modes to $\ell = 2$ modes is 0.60, which is precisely what is observed! Hence PG 1159 pulsates in $\ell = 1$ and $\ell = 2$ modes; its rotation period is 1.38 days.

The data show a complete string of 20 $\ell = 1$ modes between 430 and 840 seconds. The mean period spacing of the $\ell = 1$ modes is 21.5 seconds, and 12.5 seconds for the $\ell = 2$ modes. The observed ratio of the period spacings is 1.72; the theoretical value is 1.73. Again there is excellent agreement between theory and observation. From these period spacings equation (2) gives the mass of PG 1159 as $0.590 \pm 0.01 M_\odot$. Deviations from uniform spacing of the $m = 0$ periods are clearly apparent, and strongly suggestive of mode trapping by a subsurface composition transition. Kawaler and Bradley (1994) use the period spacing diagram to determine the thickness of the surface helium layer in PG 1159 as $3 \times 10^{-3} M_*$. The model that best fits the the mean period spacing and the observed periods had a luminosity of $220 L_\odot$, an effective temperature of $136,000 K$, $\log g = 7.4$, and a surface helium abundance of 30%. From traditional spectroscopy, Werner, Heber, & Hunger (1991) find an effective temperature of 140,000 K and $\log g = 7.0 \pm 0.5$, in excellent agreement with the results from temporal spectroscopy.

### 3.2. PG 2131+066

Another PG 1159 star examined using the WET was PG 2131+066 (hereafter PG 2131). PG 2131 was little–observed (photometrically) following its discovery as a variable star by Bond et al. (1984). Data obtained at McDonald Observatory in 1984 indicated at least 6 periodicities, but the limitations of single–site observing rendered positive period determinations impossible. The Kiel group has analyzed the spectrum of PG 2131, and report an effective temperature of about $80,000 \pm 10,000$ K, and $\log g = 7.5 \pm 0.5$ (Dreizler et al. 1995): they indicate that PG 2131 might be more helium–rich than PG 1159.

PG 2131 was the target of a WET run in September 1992. The star is significantly fainter than PG 1159; therefore the data obtained have a lower S/N level. Despite this, the coverage of PG 2131 was sufficient for Kawaler et al. (1995) to decompose the observed pulsation spectrum of PG 2131. PG 2131 pulsates in $\ell = 1$ modes with several consecutive overtones, and a period spacing of 21.7 seconds. The power spectrum shows these modes are (roughly) equally spaced triplets, giving a rotation period of 5.07 hours. PG 2131 has a mass of $0.61 \pm 0.02 M_\odot$, and there is a hint of a discontinuity in the composition near $6 \times 10^{-3} M_*$ below the surface. The best-fit $0.61 M_\odot$ models have $T_{\rm eff} = 84,000 \pm 2,000 K$ and $L = 10 \pm 4 L_\odot$.

PG 2131 shows some similarities to PG 1159, but also some important differences. The largest amplitude periods are at higher frequency than PG 1159. There are fewer modes present, however, and there is no evidence for $\ell = 2$ modes. The amplitude of the pulsations is slightly larger in PG 2131 than in PG 1159. The differences in the pulsation behavior are probably correlated to the difference in temperature between PG 1159 and PG 2131.

### 3.3. Other PG 1159 Stars

Additional WET campaigns other hot white dwarfs are currently being analyzed. PG 1707 is another naked PG 1159 star for which WET data exist and are being analyzed, though that star had a gap in coverage (corresponding to central Asia) and therefore has a significant alias problem. WET data have also been obtained for RX J2117+34, the pulsating central star of an enormous old planetary nebula. This star may prove to be extremely important, as it shows periods that are shorter than any other central star, but longer than for the naked PG 1159 stars (Vauclair et al. 1993). It may provide a direct link between the planetary nebula nuclei and hot DO white dwarfs (Appleton, Kawaler, and Eitter 1993). Other pulsating PN central stars have been actively investigated by Howard

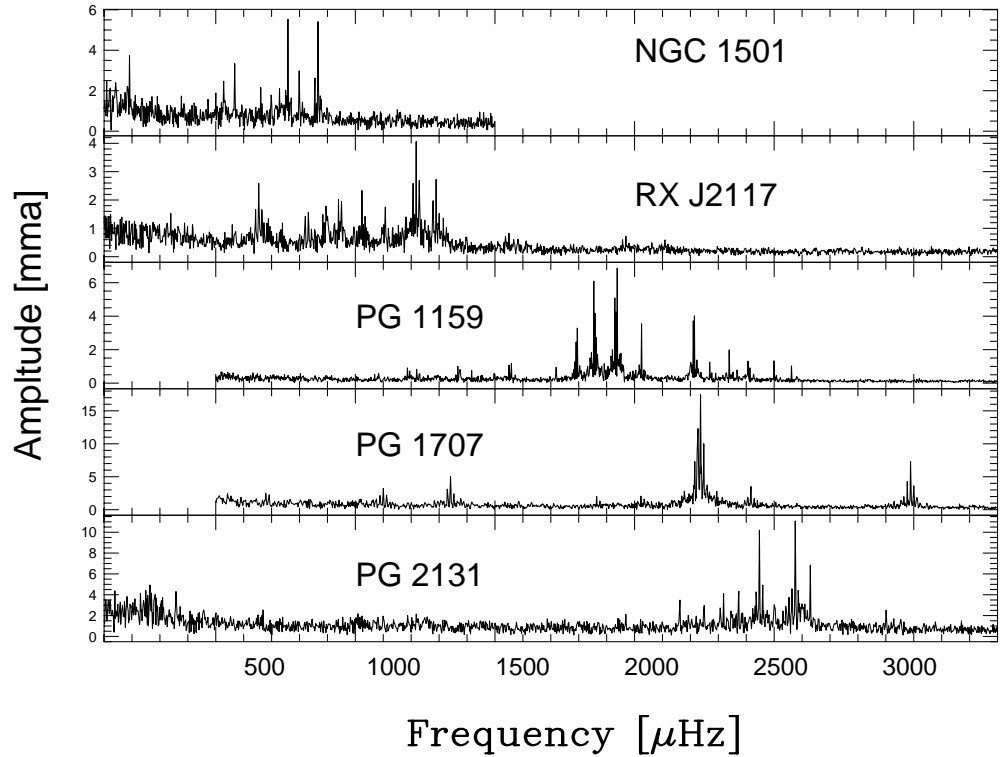

Figure 1. Power spectra of hot pulsating degenerate stars. The data for PG 1159 are from Winget et al. (1991) and for PG 2131 from Kawaler et al. (1995). The other data are from works in preparation: data for NGC 1501 were obtained by Bond et al. (1995), for RX J2117 by Vauclair et al. (1995). PG 1707 data are from an early WET run.

Bond and his collaborators. They have obtained globally coordinated CCD fast photometry of the central star of the planetary nebula NGC 1501 and Sand 3 (see Bond et al. 1993).

Figure 1 shows the power spectra of several hot white dwarfs in order of decreasing dominant period. Note the striking similarities (and differences) between these objects in terms of dominant period and distribution of peak amplitudes. It is left to the reader to derive conclusions based on the temperatures of these objects. The values of $T_{\rm eff}$, in units of $10^3$K, are 96 ± 3 for NGC 1501 (Shaw & Kaler 1985), 150 ± 15 for RX J2117 (Motch, Werner & Pakull 1993), 140 ± 10 for PG 1159, 100 ± 10 for PG 1707 (Werner, Heber & Hunger 1991), and 84 ± 2 for PG 2131 (Kawaler et al. 1995).

## 4. The DB Pulsator GD 358

Another class of pulsating white dwarfs are the pulsating DB (helium–dominated surface) white dwarfs. The prototype of this class is GD 358, which was the first member of the class to be discovered (Winget et al. 1982) following the theoretical prediction by Winget et al. (1981) that DB white dwarfs should show analogous nonradial pulsations to the ZZ Ceti stars. GD 358 has an effective temperature of about 25,000K (Thejl, Vennes, & Shipman 1991). It has an extremely complex light curve that made it an obvious target for study with the WET.

Winget et al. (1994) report on WET observations of GD 358, which have an extremely high duty cycle and clean power spectrum. They found over 180 separate pulsation frequencies in this star, including many triplets. The triplet spacing is measurably non–uniform, with the triplets at long period being systematically wider than at short period. Winget et al. (1994) interpret this as evidence of differential rotation, with the outer layers of GD 358 rotating faster than interior layers. The rotation period near the surface is 0.88 days, and 1.6 days at the deeper levels. In addition, Winget et al. (1994) find a systematic shift of the $m = 0$ peak to lower frequencies, which is the signature of magnetic fields (Jones et al. 1989). Measurement of this shift allows Winget et al. (1994) to determine the magnetic field strength of $1300 \pm 300$G. Other peaks in the power spectrum lie precisely at the sums and/or differences of the dominant "normal" peaks, demonstrating that important nonlinear effects are operating in this star.

Model analysis of GD 358 by Bradley & Winget (1994) successfully reproduces in detail the pulsation frequencies. They determine a mass of GD 358 of $0.61 \pm 0.03 M_\odot$, $T_{\text{eff}} = 24 \pm 1 \times 10^3 K$, and a luminosity of $0.050 \pm 0.012 L_\odot$. The resulting distance to GD 358 is $42 \pm 3$ pc, meaning that the pulsations of GD 358 (with a dose of theory) are a distance indicator. In principle, the same distance determination technique can be applied to the PG 1159 stars, but in those hotter stars the bolometric correction introduces a much larger uncertainty.

Perhaps the most striking result of the modeling by Bradley & Winget (1994) is that the observed periods are best matched with a surface layer of pure helium of approximately $1.2 \times 10^{-6} M_\odot$. This is three order of magnitude smaller than PG 1159. This difference in helium layer mass suggests that establishing an evolutionary connection between the PG 1159 stars and DB white dwarfs requires finely–tuned mass loss beyond the PG 1159 stage. Alternatively, these results challenge the notion of there being a direct evolutionary relationship between these objects.

Exploration of this dilemma is another example of how temporal spectroscopy of white dwarf stars drives our growing understanding of their origins and chemical evolution. Dehner (1995) and Dehner & Kawaler (1995) show that the two classes of stars *do* appear to have a direct evolutionary connection. They computed the evolution of an initial model based on the PG 1159 pulsational data, including time–dependent diffusion. As expected, diffusion causes a separation of the elements; a thickening surface layer of nearly pure helium overlays a deepening transition zone where the composition changes to the surface composition of the original model. In the temperature range inhabited by GD 358 and the pulsating DB white dwarfs, this pure helium surface layer is $\sim 10^{-5.5} M_*$ thick. The pulsation periods of the resulting model show a good fit to the WET observations. The model is very similar to the model used by Bradley and Winget (1994) to match the pulsation observations of GD 358. These results demonstrate the plausibility of a direct evolutionary path from PG 1159 stars to the much cooler DB white dwarfs.

## 5. Conclusion

The extremely sensitive data gathering and analysis techniques demonstrated by the WET collaboration have already produced data that push the limits of current theoretical tools. The link between the PG 1159 stars and DB white dwarfs is a tantalizing example of how the powerful techniques of stellar seismology probe the interiors of stars. Because of lack of space, this review has been limited to the DO and DB white dwarfs observed with the WET. Numerous studies of other types of variable stars have been made with this instrument. In these proceedings, Paul Bradley outlines some of the modeling possibilities for DA white dwarfs. J.C. Clemens (1994, 1995) has investigated the temporal spectra of a number of ZZ Ceti stars with the WET. Clemens and others find strong evidence of a fundamental similarity between the ZZ Ceti stars; the appear to have hydrogen layer masses that are entirely consistent with their origin as hydrogen–burning PN central stars that have not lost a significant amount of mass on their descent to become white dwarfs.

Clearly, temporal spectroscopy of white dwarfs, with the WET as a principal instrument, shows much more than promise. The field is extremely exciting, and becoming more exciting than ever as we probe the innards of white dwarfs and reveal the last stages of evolution of stars that were once very much like our own Sun.

## References

Appleton, P.N., Kawaler, S.D., and Eitter, J.J. 1993, AJ, 106, 1973
Bergeron, P. et al. 1995, ApJS, in press
Bond, H., Ciardullo, R. & Kawaler, S.D. 1993, ActaAstr, 43, 425
Bond, H.E., Grauer, A.D., Green, R.F., & Liebert, J. 1984, ApJ, 279, 751
Bond, H.E. et al. 1995, ApJ, in preparation
Bradley, P.A., & Winget, D.E. 1991, ApJS, 75, 463
Bradley, P.A., Winget, D.E., & Wood, M.A. 1993, ApJ, 406, 661
Bradley, P. & Winget,D. 1994, ApJ, 430, 850
Brassard, P., Fontaine, G., Wesemael, F. & Hansen, C.J. 1992a, ApJS, 80, 369
Brassard, P., Fontaine, G., Wesemael, F. & Tassoul, M. 1992b, ApJS, 81, 747
Brickhill, A.J. 1975, MNRAS, 170, 404
Clemens, J.C. 1994, Ph.D. thesis, University of Texas
Clemens, J.C. 1995, in preparation
Dehner, B.T. 1995, Ph.D. thesis, Iowa State University
Dehner, B.T., & Kawaler, S.D. 1995, ApJ, in press
Dreizler, S. et al. 1995, in 9th European Workshop on White Dwarf Stars, ed. D. Koester & K. Werner (Dordrecht: Kluwer), in press
Jones, P.W., Pesnell, W.D., Hansen, C.J., & Kawaler, S.D. 1989, ApJ, 336, 403
Kawaler,S. D. 1987, in Stellar Pulsation: A Memorial to John P. Cox, eds. A. Cox, W. Sparks, & S. Starrfield, (Berlin: Springer–Verlag), 367

Kawaler, S.D. 1988, in IAU Symposium 123, Advances of Helio- and Asteroseismology, ed. J. Christensen–Dalsgaard & S. Frandsen (Dordrecht: Reidel), 329

Kawaler, S.D. & Hansen, C.J. 1989, in I.A.U. Colloquium #114: White Dwarfs, ed. G. Wegner, (Berlin: Springer–Verlag), 97.

Kawaler, S.D., & Weiss, P. 1990, in Proc. of the Oji International Seminar, Progress of Seismology of the Sun and Stars, ed. Osaki, Y.,& Shibihashi, H., (Berlin: Springer), 431

Kawaler, S.D. 1991, in Confrontation Between Stellar Pulsation and Evolution, ed C. Cacciari & G. Clementini, (Provo: PASP), 494

Kawaler, S.D. & Bradley, P.A. 1994, ApJ, 427, 415

Kawaler, S.D. et al. 1995, ApJ, in press

McGraw, J.T., Starrfield, S.G., Liebert, J., & Green, R.F. 1979, in IAU Colloq. 53, White Dwarfs and Variable Degenerate Stars, ed. H.M. Van Horn & V. Weidemann, (Rochester: Univ. of Rochester), 377

Motch, C., Werner, K, & Pakull, M.W. 1993, A&A, 268, 561

Nather, R.E., Winget, D.E., Clemens, J.C., Hansen, C.J., & Hine, B.P. 1990, ApJ, 361, 309

Shaw, R. & Kaler, J. 1985, ApJ, 295, 537

Tassoul, M., Fontaine, G., & Winget, D.E. 1990, ApJS, 72, 335

Thejl, P., Vennes, S., & Shipman, H.L. 1991, ApJ, 370 355

Vauclair, G., Belmonte, J.A., Pfeiffer, B., Chevreton, M., Dolez, N., & Motch, C. 1993, A&A, 267, L35

Vauclair, G., et al. 1995, ApJ, in preparation

Werner, K., Heber, U., & Hunger K. 1991, A&A, 244, 437

Winget, D.E. 1981, Ph.D. thesis, University of Rochester

Winget, D.E., Van Horn, H.M., & Hansen, C.J. 1981, ApJ, 245, L33.

Winget, D.E., Robinson, E.L., Nather, R.E., & Fontaine, G. 1982, ApJ, 262, L11

Winget, D.E., Van Horn, H.M., Tassoul, M., Hansen, C., & Fontaine, G. 1983, ApJ, 268, L33

Winget, D.E. 1988, in IAU Symposium 123, Advances in Helio- and Asteroseismology, ed. J. Christensen–Dalsgaard & S. Frandsen, (Dordrecht: Reidel), 305

Winget, D.E., et al. 1991, ApJ, 378, 326

Winget, D.E., et al. 1994, ApJ, 430, 839.

Wood, M.A. 1992, ApJ, 386, 539